\documentclass[final,5p,times,twocolumn]{elsarticle}

\usepackage{amsmath}
\usepackage{graphicx} 
\usepackage{tabularx}
\usepackage[euler]{textgreek}
\usepackage{upgreek}
\usepackage{comment}
\usepackage{enumitem}
\usepackage{subcaption}

\usepackage{lineno,hyperref}

\begin{document}

\title{Detector Requirements, Design, and Technologies for the FCC-ee
Higgs, electroweak, and top factory}


\author{Mogens Dam}
\address{Niels Bohr Institute, Copenhagen University, Jagtvej 155, 2200 Copenhagen, Denmark}
\ead{dam@nbi.dk}

\date{April 2025}

\begin{abstract}
The proposed high-luminosity, circular electron-positron collider,
\mbox{FCC-ee}, 
provides unparalleled opportunities for precise exploration of Higgs, electroweak, top, flavour, and beyond standard model physics. Very advanced detector systems are required to fully exploit this diverse
physics programme. Key requirements include excellent resolutions on the
measurement of momentum, energy, and impact parameters; exquisite particle
identification capabilities over a wide momentum range including photon/$\pi^0$
separation; sensitivity to far-displaced vertices in the tracking (and
possibly also the calorimeter) volume; and very precise absolute and relative
normalisation. This note presents an overview of detector requirements
and the status of detector design efforts.
\end{abstract}

\maketitle

\section{Introduction}

Electron-positron collisions at the Future Circular Collider (FCC-ee) provide unparalleled opportunities for Higgs, electroweak (EW), flavour and beyond standard model (BSM) physics.
High-energy $\mathrm{e^+e^-}$ collisions provide a clean, controlled, and well-defined environment avoiding the complications
present in proton-proton collisions. Unlike proton-proton collisions, where parton distribution functions, multiparton interactions and ``beam remnants'' complicate the understanding, $\mathrm{e^+e^-}$ annihilations occur at a fixed centre-of-mass energy with precisely known elementary kinematics. 

The development of detector systems to fully exploit the vast FCC-ee physics potential is an active area of ongoing study. Despite the different conditions, much can be learned from the LHC detectors and their high-luminosity upgrades. Closer in terms of experimental conditions are studies over the past decades of detectors for linear colliders. Important differences here arise from the (much) higher luminosity performance of FCC-ee and the resulting wider physics scope.

More detailed information on the subject of this note can be found in the recently published FCC Feasibility Study Report~\cite{Benedikt:2025hsi}.

\section{The FCC-ee, Circular \texorpdfstring{e$^+$e$^-$}{e+e-} Higgs, Electroweak and Top Factory}

With its high luminosity, its clean experimental conditions, its multiple interaction regions, and a range of energies that cover the four heaviest particles of the Standard Model (SM), FCC-ee offers a uniquely broad and powerful physics exploration programme as a Higgs, electroweak, QCD, flavour, and top factory, with high potential for 
discoveries. 
The baseline plan considers operation with detectors at four interaction points (IPs), at e$^+$e$^-$ centre-of-mass energies ranging from the Z pole, via the WW threshold and the ZH production maximum, up to and just above the $\mathrm{t\bar{t}}$ threshold. 
%
The current values for the luminosities expected at these energies and the envisioned 15-year experimental programme are summarised in Table~\ref{tab:seqbaseline}, together with the number of events expected at each energy. 


\begin{table*}[ht]
\renewcommand{\arraystretch}{1.05}
\centering
\caption{The FCC-ee operation model with four interaction points, showing the centre-of-mass energies, design instantaneous luminosities for each IP, and integrated luminosity per year summed over 4~IPs.  
The last two rows indicate the total integrated luminosity and number of events expected to be produced in the four detectors. 
The number of WW events includes all $\sqrt{s}$ values from 157\,GeV up.
}
\label{tab:seqbaseline}
\begin{tabular}{lccccccc} \\[-3mm]
\hline 
Working point & Z pole & WW thresh.\ & ZH & \multicolumn{2}{c}{$\mathrm{t\bar{t}}$} \\ \hline
$\sqrt{s}$ {(GeV)} & 88, 91, 94 & 157, 163 & 240 & 340--350 & 365 \\ 
Lumi/IP {($10^{34}$\,cm$^{-2}$s$^{-1}$)} & 140 & 20 & 7.5 & 1.8 & 1.4 \\ 
Lumi/year {(ab$^{-1}$)} & 68 & 9.6 & 3.6 & 0.83 & 0.67 \\ 
Run time {(year)} & 4 & 2 & 3 & 1 & 4 \\ 
Integrated lumi.\ {(ab$^{-1}$)} & 205 & 19.2 & 10.8 & 0.42 & 2.70 \\ \hline
 &  &  & $2.2 \times 10^6$ ZH & \multicolumn{2}{c}{$2 \times 10^6$ $\mathrm{t\bar{t}}$} \\
Number of events &  $6 \times 10^{12}$ Z & $2.4 \times 10^8$ WW & $+$ & \multicolumn{2}{c}{$+\,370$k ZH} \\
 &  &  & 65k WW $\rightarrow$ H  & \multicolumn{2}{c}{$+\,92$k WW $\rightarrow$ H} \\ \hline
\end{tabular} 
\end{table*}

\section{Physics Programme and Detector Requirements}

The main challenge for the design of detectors for \mbox{FCC-ee} arises from the richness of the physics programme and the enormous event samples, especially at the Z-pole run. Matching the experimental accuracy to the statistical precision and the detector configuration to the variety of physics channels,
leads to a wide range of challenging performance requirements. 
Schematically, the endeavour can be divided into four main categories with corresponding (often overlapping) sets of requirements: \emph{i}) the Higgs factory programme; \emph{ii}) the precision electroweak programme; \emph{iii}) the flavour programme; and \emph{iv}) the direct search for feebly interacting BSM
particles. The situation is outlined in Table~\ref{tab:requirements} where the left column summarises these four main categories,
and the right column lists the corresponding key detector requirements.

\begin{table*}[ht]
\renewcommand{\arraystretch}{1.05}
\centering
\caption{Schematic overview of FCC-ee Detector Requirements.}
\label{tab:requirements}
\begin{tabularx}{\textwidth}{XX} \\[-3mm]
\hline 
\textbf{Physics programme} & \textbf{Detector Requirements} \\ \hline
\textbf{Higgs Factory}
\begin{itemize}[noitemsep,topsep=0pt,align=left,itemindent=0mm,labelsep=3pt,leftmargin=*]
    \item[-] At $\sqrt{s} = 240$\,GeV and 365\,GeV, collect 2.6M HZ and 150k WW $\rightarrow$ H events
    \item[-] Higgs properties: mass and width, couplings to fermions and bosons, self-coupling measurement via loop diagrams
\end{itemize}
&
\begin{itemize}[noitemsep,topsep=0pt,align=left,itemindent=0mm,labelsep=3pt,leftmargin=*]
    \item[-] Momentum resolution $\delta(p_\mathrm{T})/p_\mathrm{T} \simeq 10^{-3}$ at $p_\mathrm{T} \simeq 50$\,GeV 
    \item[-] Jet energy 
    resol $\delta(E)/E \simeq$ 3--4\,\% for Z/W/H 
    classification
    \item[-] Superior impact parameter resolution for b, c tagging
    \item[-] Charged hadron PID for s tagging
\end{itemize} \\[-4mm] \hline
\textbf{Precision electroweak}
\begin{itemize}[noitemsep,topsep=0pt,align=left,itemindent=0mm,labelsep=3pt,leftmargin=*]
    \item[-] Collect $6\times 10^{12}$ Z and $2.4 \times 10^8$ WW events:
    \item[] Factor $\sim$500 improvement of statistical precision on EWPOs \newline
    \phantom{M} $m_\mathrm{Z}$, $\Gamma_\mathrm{Z}$, $\Gamma_\mathrm{inv}$, $\sin^2\theta_\mathrm{W}$, $R_\ell$, $R_\mathrm{b}$, $m_\mathrm{W}$, $\Gamma_\mathrm{W}$, \dots
    \item[-] Collect $2\times 10^6$ $\mathrm{t\bar{t}}$ events: 
    $m_\mathrm{top}$, $\Gamma_\mathrm{top}$, EW and Higgs couplings
    \item[-] Indirect sensitivity to new physics up to several tens of TeV
\end{itemize} 
&
\begin{itemize}[noitemsep,topsep=0pt,align=left,itemindent=0mm,labelsep=3pt,leftmargin=*]
    \item[-] Absolute normalisation to 10$^{-4}$ or better
    \item[-] Relative normalisation between channels and between energy-scan points to $10^{-5}$
    \item[-] Track angular resolution to 0.1\,mrad
    \item[-] Stability of B field (momentum scale) to $10^{-6}$ 
\end{itemize} \\[-4mm] \hline
\textbf{Flavour}
\begin{itemize}[noitemsep,topsep=0pt,align=left,itemindent=0mm,labelsep=3pt,leftmargin=*]
    \item[-] From Z decays: $10^{12}$ $\mathrm{b\bar{b}}$, $\mathrm{c\bar{c}}$, $2\times 10^{11}$ $\uptau^+\uptau^-$, clean and boosted
    \item[-] Heavy quark decays: FCNC searches, CKM matrix, CP measurements
    \item[-] \texttau{} decays: LFV searches, lepton universality tests
\end{itemize}
&
\begin{itemize}[noitemsep,topsep=0pt,align=left,itemindent=0mm,labelsep=3pt,leftmargin=*]
    \item[-] Superior impact parameter resolution
    \item[-] Precise identification and measurement of secondary vertices
    \item[-] ECAL resolution of few \%/$\sqrt{E}$; excellent $\uppi^0/\gamma$ separation
    \item[-] PID: charged hadron identification 
    over wide momentum range
\end{itemize} \\[-4mm] \hline
\textbf{BSM particle searches}
\begin{itemize}[noitemsep,topsep=0pt,align=left,itemindent=0mm,labelsep=3pt,leftmargin=*]
    \item[-] High sensitivity direct searches for feebly interacting particles with masses below $m_\mathrm{Z}$
    \item[-] Axion-like particles, dark photons, Heavy Neutral Leptons
    \item[-] Long-lifetime signatures, LLPs
\end{itemize}
&
\begin{itemize}[noitemsep,topsep=0pt,align=left,itemindent=0mm,labelsep=3pt,leftmargin=*]
    \item[-] Hermeticity
    \item[-] Precise timing
    \item[-] Sensitivity to (far) detached vertices (mm to m range)
    \item[] - Tracking: more layers, ``continous tracking''
    \item[] - Calorimetry: granularity, tracking capabilities
\end{itemize} \\[-4mm] \hline
\end{tabularx} 
\end{table*}

\subsection{Higgs programme}

The properties of the Higgs boson will be studied through datasets collected at 240\,GeV and 340--365\,GeV, allowing for both the Higgs-strahlung (e$^+$e$^-$ $\rightarrow$ HZ) and the WW-fusion (e$^+$e$^-$ $\rightarrow \text{W}^+ \text{W}^- \upnu_\mathrm{e}\bar{\upnu}_\mathrm{e} \rightarrow$ H$\upnu_\mathrm{e}\bar{\upnu}_\mathrm{e}$) processes to be exploited, with the higher energies also serving as a laboratory for top quark physics. The separation of HZ events
from backgrounds is based on the 
reconstruction of the Z mass and  the \textit{Z recoil mass} (the mass of the system against which the Z recoils). For the cleanest signatures where the Z decays leptonically ($\text{Z} \rightarrow$ e$^+$e$^-$, \textmu$^+$\textmu$^-$), reconstruction is dependent on an excellent momentum resolution.
For the dominant Z $\rightarrow$ jj decay modes, an excellent jet-energy resolution is crucial, as it also is for the reconstruction of the Higgs boson itself that also predominantly decays to hadronic final states.  
The precise measurement of the Higgs couplings to b, c, and s quarks and to gluons depends on the determination of the jet origin in \mbox{H $\rightarrow$ jj} decays. Crucial ingredients are a superior impact parameter resolution for b and c quark tagging, and charged particle identification (PID) for the tagging of the  rare \mbox{($\sim$ 0.1\%) H $\rightarrow \mathrm{s\bar{s}}$} decays.





\subsection{Precision electroweak programme}

The exceptional luminosity performance
at the Z-resonance peak and 
at the WW threshold 
allows for measurement of electroweak precision observables (EWPOs) with unprecedented precision.
Compared to LEP, statistical uncertainties will decrease by more than two orders of magnitude. 
Matching this precision with a comparable systematic accuracy relies
on accurate knowledge of beam parameters, such as centre-of-mass energy, energy spread, and luminosity, and on a detailed understanding of detector acceptance and reconstruction efficiencies. While
the beam energy will be measured with the resonant depolarization technique~\cite{Blondel:2019jmp}, luminosity will be determined via the precise measurement of QED reference processes. Two complementary processes are available: small-angle Bhabha scattering, $\mathrm{e^+e^- \rightarrow e^+e^-}$, observed in dedicated luminosity monitors, and large-angle diphoton production, $\mathrm{e^+e^- \rightarrow \upgamma\upgamma}$, observed in the main detector.
An ambitious precision goal of 10$^{-4}$ ($2 \times 10^{-5}$) on the absolute luminosity measurement from the Bhabha scattering (diphoton) method translates into a required precision of about only 1\,\textmu{}rad (8\,\textmu{}rad) on the angle defining the acceptance of the luminosity monitor (central calorimetry).

\subsection{Flavour programme}

With about $10^{12}$ Z decays to each of the quark flavours, $\mathrm{b\bar{b}}$ and $\mathrm{c\bar{c}}$, and $2\times 10^{11}$ decays to $\uptau^+\uptau^-$, the Z-pole run forms the cornerstone of the FCC-ee flavour programme. Event statistics exceeds that projected for Belle~II by about an order of magnitude, and decays are strongly boosted, an advantage for most studies.  An important detector requirement is the precise measurement of the impact parameter of charged particles, the basis for the reconstruction of secondary (and tertiary) vertices for decay mode identification and lifetime measurements. Charged hadron identification, in particular \textpi/K separation, over a wide momentum range, is an important requirement, as is a high-resolution measurement of electromagnetic showers, both in terms of energy and position, for the measurement of final states with photons and \textpi$^0$s.

\subsection{BSM particle searches}

The large Z sample is the main engine that drives also searches for feebly interacting BSM particles. Candidate particles are numerous with prominent examples being Heavy Neutral Leptons (N), axion-like particles (a), and dark photons (A$^\prime$). Possible signatures include mono-jets 
(e.g., $\text{Z} \rightarrow \upnu \text{N},\ \text{N} \rightarrow \text{jet}$),
long-lived particles (LLPs) giving rise to (far) detached vertices,
three-photon final states 
(e.g., from $\text{Z} \rightarrow \text{a}\upgamma,\ \text{a} \rightarrow \upgamma\upgamma$), 
and, in general, final states where a given combination of particles give rise to a mass peak
(e.g., $\text{Z} \rightarrow \text{A}^\prime\upgamma,\ \text{A}^\prime \rightarrow \upmu^+\upmu^-)$. 
Crucial for searches for exotic signatures is hermeticity of the detector system. For searches for LLPs, the ability to identify decay vertices in a large detector volume, possibly including a fine-grained calorimeter system, is essential. The LLPs may be heavy and move slowly, and it is vital that any trigger and data acquisition system will be open for the recording of ``out-of-time'' signals.

\section{Experimental situation and challenges}

The exceptional luminosity performance of the 91-km circumference FCC-ee is achieved via a design based on the crab-waist collision scheme, with nano-beams at the interaction point (IP), and a large (30\,mrad) horizontal crossing angle. 
The collider has separate rings for the electron and positron beams, and the design is optimized for a 50\,MW synchrotron radiation energy loss per beam across the operating energy range. To compensate for the short beam lifetimes of only a few minutes, beam particles are being continuously replenished via top-up injection from a booster ring. More details can be found in \mbox{Ref.\ \cite{Benedikt:2928793}}.

\subsection{Machine Detector Interface}

\begin{figure}[ht]
\centering
\includegraphics[width=\linewidth]{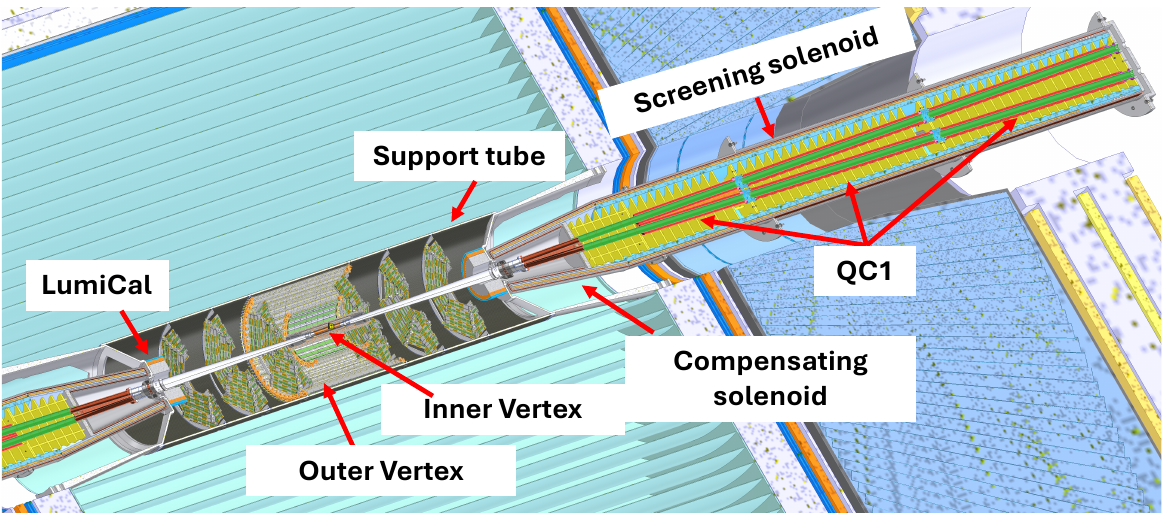}
\caption{Layout of the interaction region. 
The final focus quadrupole (QC1) is shown with the screening and compensating solenoids.
A support tube allows the integration of the luminosity calorimeter (LumiCal) and the vertex detector.}
\label{fig:IR}
\end{figure}

The FCC-ee Machine Detector Interface (MDI) has a compact and complex design \cite{Benedikt:2025hsi,boscolo_2025_p44x1-18z28}. 
The distance between the IP and the face of the superconducting final focus quadrupole ($\ell^*$) is 2.2\,m, well inside the detector volume. 
To preserve the low beam emittance and counteract the beam rotation and deflection that would be caused by the crossing of the beam and the detector magnetic field under an angle of 15\,mrad, the detector field is limited to 2\,T (at the Z-pole energy), and a compensating solenoid delivering a magnetic field of $-5$\,T is placed starting at 1.23\,m from the IP, cancelling out the longitudinal magnetic field integral along the $z$ axis from the last focussing quadrupole to the IP. Consequently, the front face of the luminosity calorimeter (LumiCal), placed in front of the compensating solenoid, is only about 1.1\,m from the IP. The layout of the interaction region
is shown in \mbox{Fig.\ \ref{fig:IR}}.

\subsection{Bunch structure and data rates}

At the four FCC-ee energy points, bunch crossings are delivered at rates of 40, 6, 1, and 0.2\,MHz, respectively. Crossings occur in a continuous mode without the use of bunch trains, as is known from linear colliders. As a consequence, counter to the situation at linear colliders where power-pulsing schemes have been developed, the front-end electronics has to be permanently powered, and given the much higher data rates, the instantaneous power (and cooling) demand is also higher.

The very high luminosity of the Z-pole run places particularly stringent requirements on the detectors and their read-out systems. With a peak cross section of 35\,nb, the rate of Z decays will be about 50\,kHz. In addition comes a similar rate of small-angle Bhabha scattering events observed in the LumiCals and a non-negligible rate of two-photon physics processes ($\upgamma\upgamma \rightarrow \text{hadrons}$). Overall, the detectors have to be able to handle event rates of order 200\,kHz.

\section{Detector Concepts}

Four detector concepts are, so far, under study for FCC-ee. 
These concepts may or may not become proposals for future FCC-ee detectors, as the result of ongoing and future 
performance studies and R\&D efforts may motivate new developments and change the overall landscape significantly. 
\begin{figure*}[htb]
\centering
\begin{subfigure}[t]{0.26\textwidth}
            \raisebox{.04\textwidth}{%
    \includegraphics[width=\textwidth]{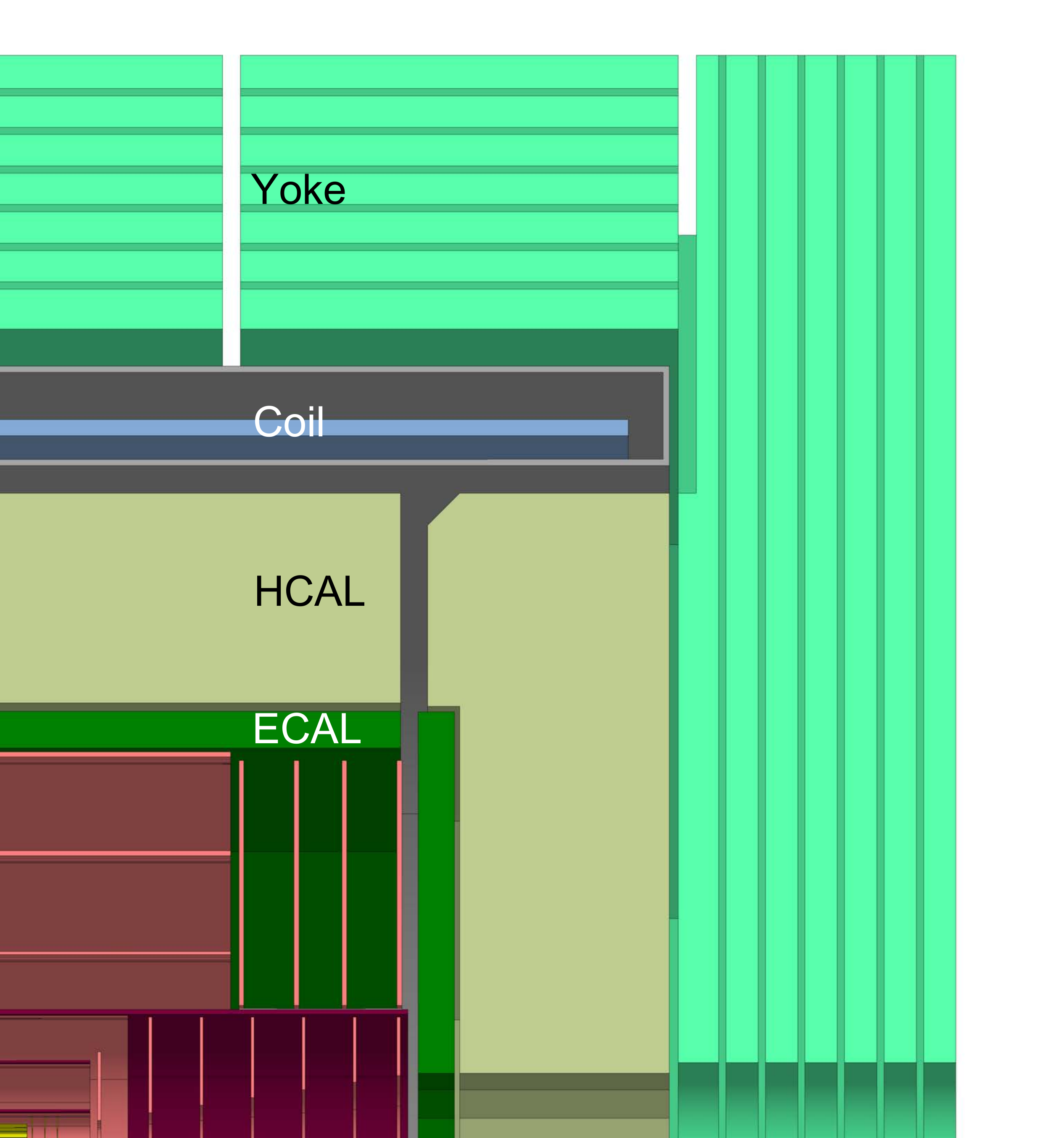}
    }
    \caption{CLD}
    \label{fig:CLD}
\end{subfigure}
\hfill
\begin{subfigure}[t]{0.36\textwidth}
    \includegraphics[width=\textwidth]{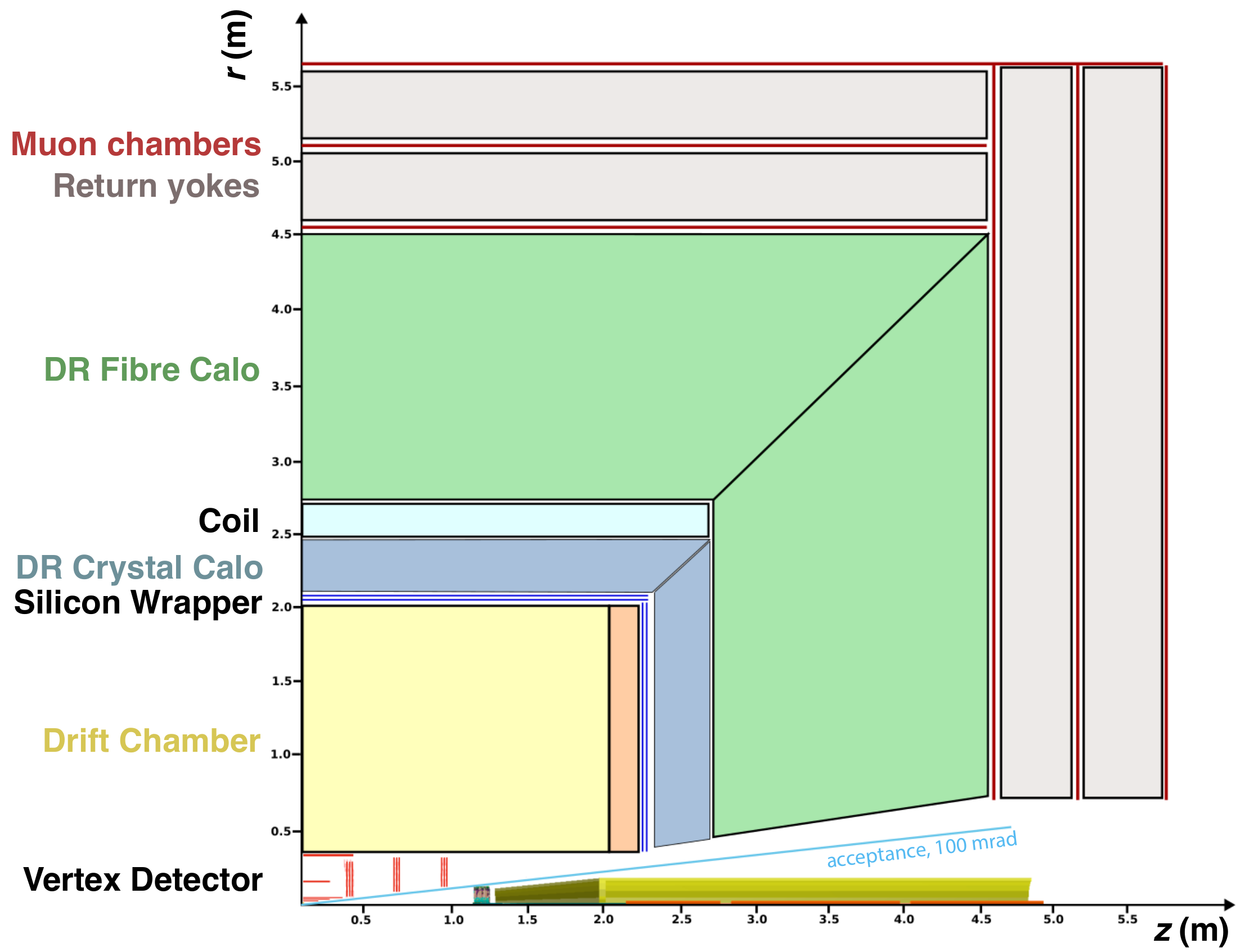}
    \caption{IDEA}
    \label{fig:IDEA}
\end{subfigure}
\hfill
\begin{subfigure}[t]{0.35\textwidth}
    \includegraphics[width=\textwidth]{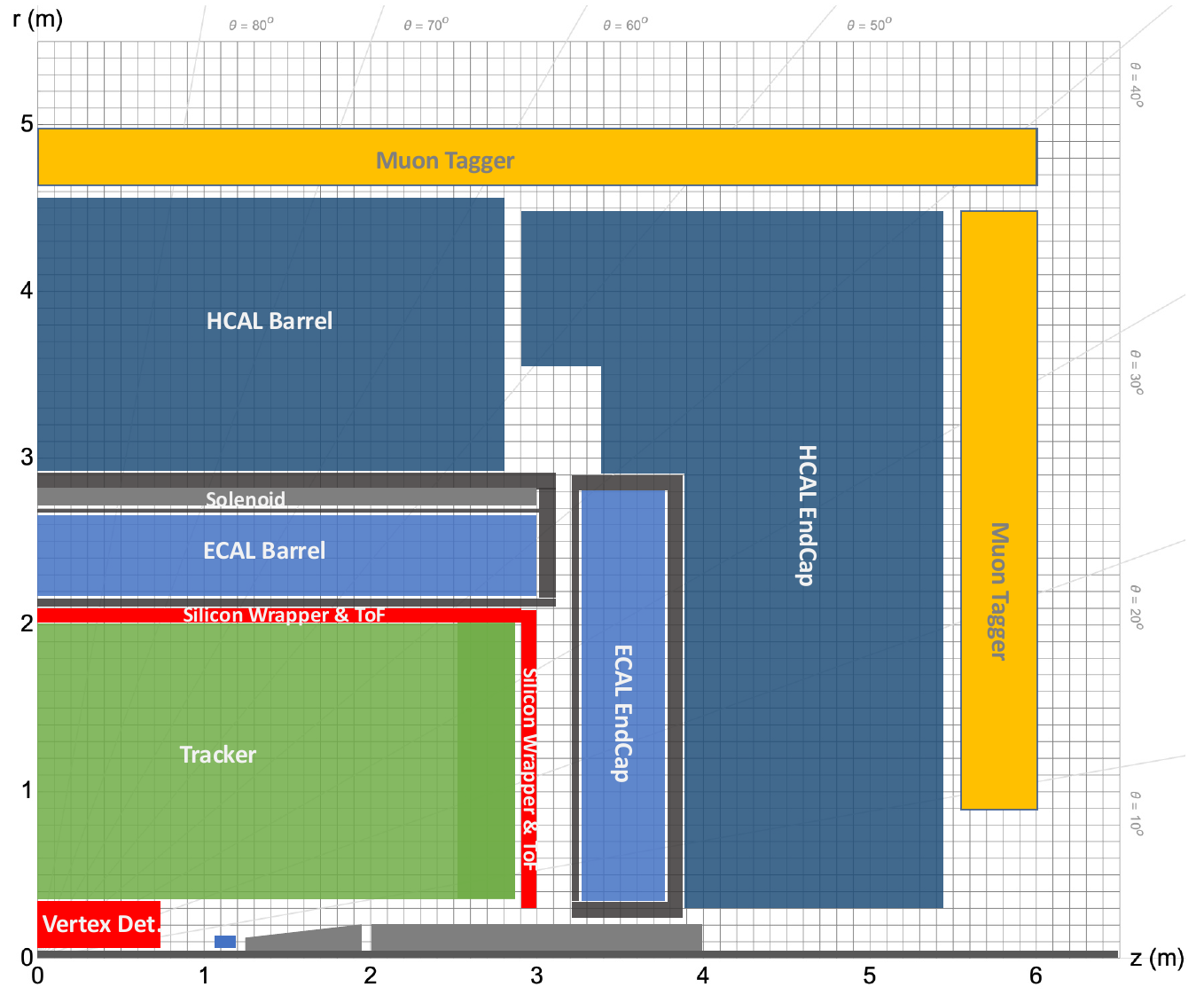}
    \caption{Allegro}
    \label{fig:Allegro}
\end{subfigure}
\label{fig:Concepts}
\caption{Longitudinal cross section of the top right quadrant of FCC-ee detector concepts. ILD has many similarities with CLD, but has a TPC as its main tracker.}
\end{figure*}
The concepts under study typically have a height of 11\,m and a similar length. To compensate for their relatively weak, 2\,T magnetic field, tracking volumes are rather large with a radius of about 2.1\,m and a similar half-length. 

The four concepts, described in more detail in \mbox{Ref.\ \cite{Benedikt:2025hsi}}, are:
\begin{description}
\item[CLD] The CLD detector concept~\cite{Bacchetta:2019fmz}, illustrated in Fig.\ \ref{fig:CLD}, is an adaptation to FCC-ee experimental conditions of the CLIC detector model~\cite{Linssen:2012hp,CLICdp:2018vnx}, itself developed from the ILD and SiD concepts~\cite{Behnke:2013lya}, originally designed for ILC. The detector features a silicon pixel vertex detector and a silicon tracker, followed by a very-high-granularity calorimeter system in form of a silicon-tungsten electromagnetic calorimeter (ECAL) and a scintillator-steel hadron calorimeter (HCAL). A superconducting solenoid, surrounding the calorimeter system, provides a 2\,T magnetic field, and a steel yoke interleaved with resistive plate muon chambers (RPC) closes the field. 
%
\item[IDEA] The IDEA detector concept~\cite{IDEAStudyGroup:2025gbt}, illustrated in Fig.\ \ref{fig:IDEA}, has been developed specifically for FCC-ee.
It comprises a silicon pixel vertex detector, a large-volume, 
extremely-light, short-drift wire chamber, for central tracking and particle identification via ionisation measurement, and a layer of silicon micro-strip detectors for improved momentum measurement. The calorimeter system is based on the dual-readout technique: a finely segmented high-resolution crystal ECAL, and a 
scintillating/Cherenkov fibre HCAL. A thin, low-mass superconducting solenoid providing a 2\,T field is positioned between the two calorimeter compartments, and a muon system based on \textmu{}-RWELL chambers is placed in the magnet return yoke.

\item[ALLEGRO]
The ALLEGRO detector concept is relatively recent and under rapid development. Its design, illustrated in Fig.\ \ref{fig:Allegro}, is articulated around a high-granularity noble-liquid ECAL.
The tracking system consists of a pixel vertex detector and a main tracer that remains to be identified; both full-silicon and gaseous options (wire chamber or straw chamber) are under study. In case of a gaseous tracker, it would be surrounded by a Si micro-strip layer.
A thin, lightweight coil follows the ECAL, with which it shares cryostat. The detector is completed with a scintillator-steel HCAL and muon chambers inside the magnet return yoke.

\item[ILD] The ILD detector concept~\cite{Behnke:2013lya}, originally developed for ILC, bears many similarities with CLD. The main difference and the signature element of ILD is a large-volume time projection chamber (TPC). While introducing a minimum of material, the TPC provides precise and continuous tracking along with PID capabilities via ionisation measurement. While studies show that a TPC is well suited for operation at all energies above the Z pole, additional studies are needed to understand whether the long drift distances of this detector are compatible with Z-pole operation.

\end{description}


\section{Detector Components and Technologies}

This section presents an overview of the main components of FCC-ee detector systems, the requirements, and proposed technologies. Again, more details can be found in \mbox{Ref.\ \cite{Benedikt:2025hsi}}.

\subsection{Vertex Detector}

The vertex detector is driving the measurement of track impact parameters and thus the reconstruction of detached vertices, important for the identification and lifetime measurement of long-lived particles such as heavy quarks and tau leptons. The impact parameter resolution is often parameterised as
\begin{equation}
    \sigma(d_0) = a \oplus \frac{b}{p \sin^{3/2}\theta},
\end{equation}
where $\theta$ is the track polar angle, and the symbol $\oplus$ signifies the square root of the quadratic sum.
Here, the asymptotic term, $a$, is driven by the single hit resolution, while the second term represents the contribution from multiple scattering in the material of the beam pipe and the vertex detector layers. Providing a first measurement point as close as possible to the IP is crucial. 
%
The FCC-ee beam pipe has an inner radius of 10\,mm, and its liquid-cooled, double-layer structure has a thickness of 1.7\,mm corresponding to 0.61\% of a radiation length ($X_0$).


The technologies available for vertex detector design -- pixel sensors, readout electronics, mechanical support structures, and cooling --  are rapidly evolving.
The sensor technology of choice 
is CMOS MAPS (Monolithic Active Pixel Sensors), where the sensitive volume and the readout circuitry are combined in a single piece of silicon. Such sensors can be thinned down to a thickness of 50\,\textmu{}m or less, providing a very transparent solution. Ultimately, these thin sensors can be bent into cylindrical layers around the beam pipe requiring very little support. 

\begin{figure}[ht]
\centering
\includegraphics[trim=0 14mm 0 0,width=\linewidth]{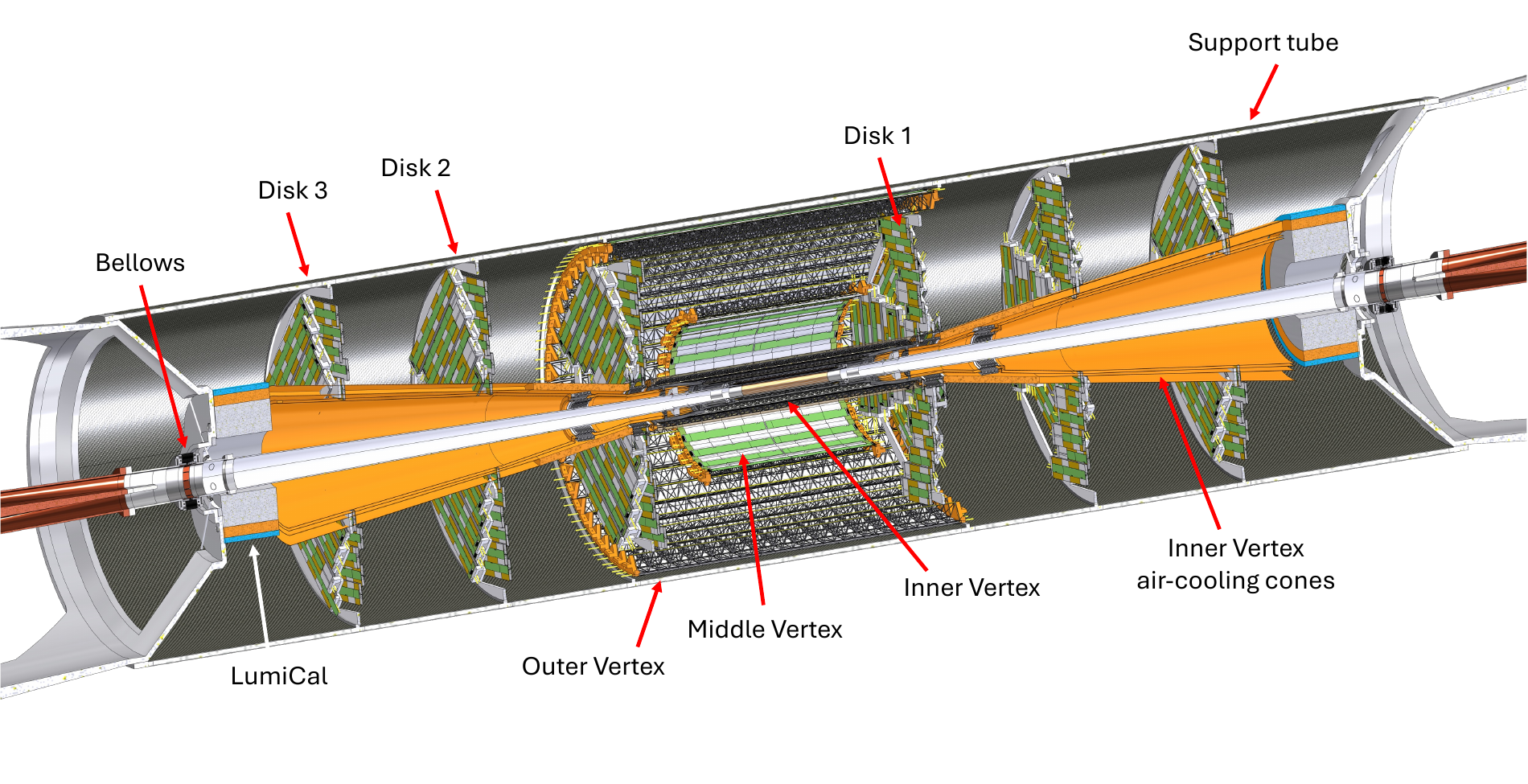}
\caption{Section of the MDI for about $\pm 1.3$\,m from the IP for the IDEA detector. Inside a support tube of about 32\,cm radius is seen the vertex detector (three inner, one middle, and one outer layers; three disks per side) and the LumiCals.}
\label{fig:vtx}
\end{figure}

\begin{figure}[ht]
\centering
\includegraphics[width=0.7\linewidth]{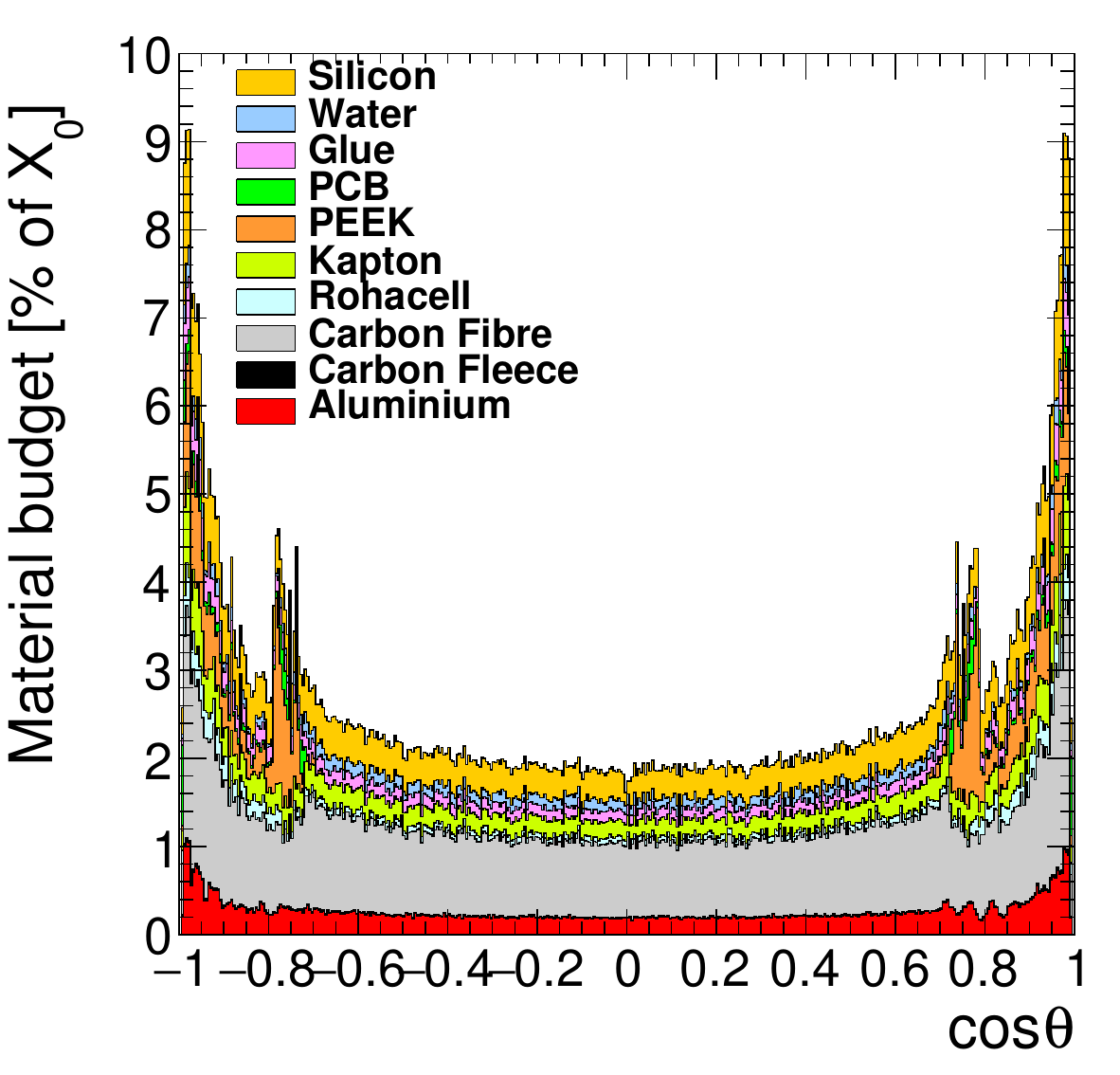}
\caption{Material budget 
for the vertex detector considered for ALLEGRO and IDEA as a function of the cosine of the polar angle.}
\label{fig:VTX_X0}
\end{figure}

In an engineering-level study, considered for 
ALLEGRO and IDEA, 
a vertex detector design, integrated into the complex MDI layout, has been developed.
A light carbon-fibre support tube with a radius of 32\,cm surrounds the vertex detector as well as the LumiCals. 
The vertex detector layout, illustrated in \mbox{Fig .\ \ref{fig:vtx}}, includes a barrel section with three inner, one middle, and one outer layers, and three disks in each direction.
For all layers, the sensors are 50\,\textmu{}m thick, planar MAPS.
The inner barrel layers have $25 \times 25\,\text{\textmu{}m}^2$ pixels that, via pulse-height information and charge sharing, provide a single-point resolution of $\sim$3\,\textmu{}m. 
Sensors 
are mounted on a light carbon-fibre support structure, the assembly contributing about 0.25\% $X_0$. An estimated power consumption of 50\,mW\,cm$^{-2}$ allows the use of air cooling. 
%
The middle and outer barrel layers and the disks have sensors with $150 \times 50\,\text{\textmu{}m}^2$ pixels. 
Sensor modules are
mounted on carbon-fibre support structures
and are cooled by water that circulates in thin polyimide tubes. The complete vertex detector is very transparent with less than 2.5\% $X_0$ of material at normal incidence, as illustrated in Fig.\ \ref{fig:VTX_X0}.

An alternative, ultra-light layout for the inner vertex detector is also being explored. The layout is based on a concept similar to the ALICE ITS3~\cite{Mager:2747243} curved sensor technology that facilitates a self-supporting structure with essentially no material other than that of the sensors. The technology uses a stitching technique to form wafer-scale sensors from multiple repeated sensor units. With four layers made from half-cylindrical sensors, 
%
the material budget is reduced by $\sim$60\% compared to the baseline design. 

\subsection{Tracking System}

The efficient reconstruction of charged-particle trajectories and the precise measurement of momentum
are crucial requirements for all physics analyses. The (transverse) momentum resolution is often parameterised as
\begin{equation}
    \frac{\sigma(p_\mathrm{T})}{p_\mathrm{T}} = (a \cdot p_\mathrm{T}) \oplus b,
\end{equation}
where 
the first, asymptotic term is driven by the detector resolution, while the second term represents the contribution from multiple scattering.
Two main technologies are being considered for the main tracking system: \emph{i}) solid state detectors based on high-resolution Si sensors, and \emph{ii}) gaseous trackers. In case of a gaseous tracker, this would be surrounded by a Si-based system (a \emph{wrapper}) to provide an additional precise space point for improved momentum resolution. 

\begin{figure}[ht]
\centering
\includegraphics[width=0.75\linewidth]{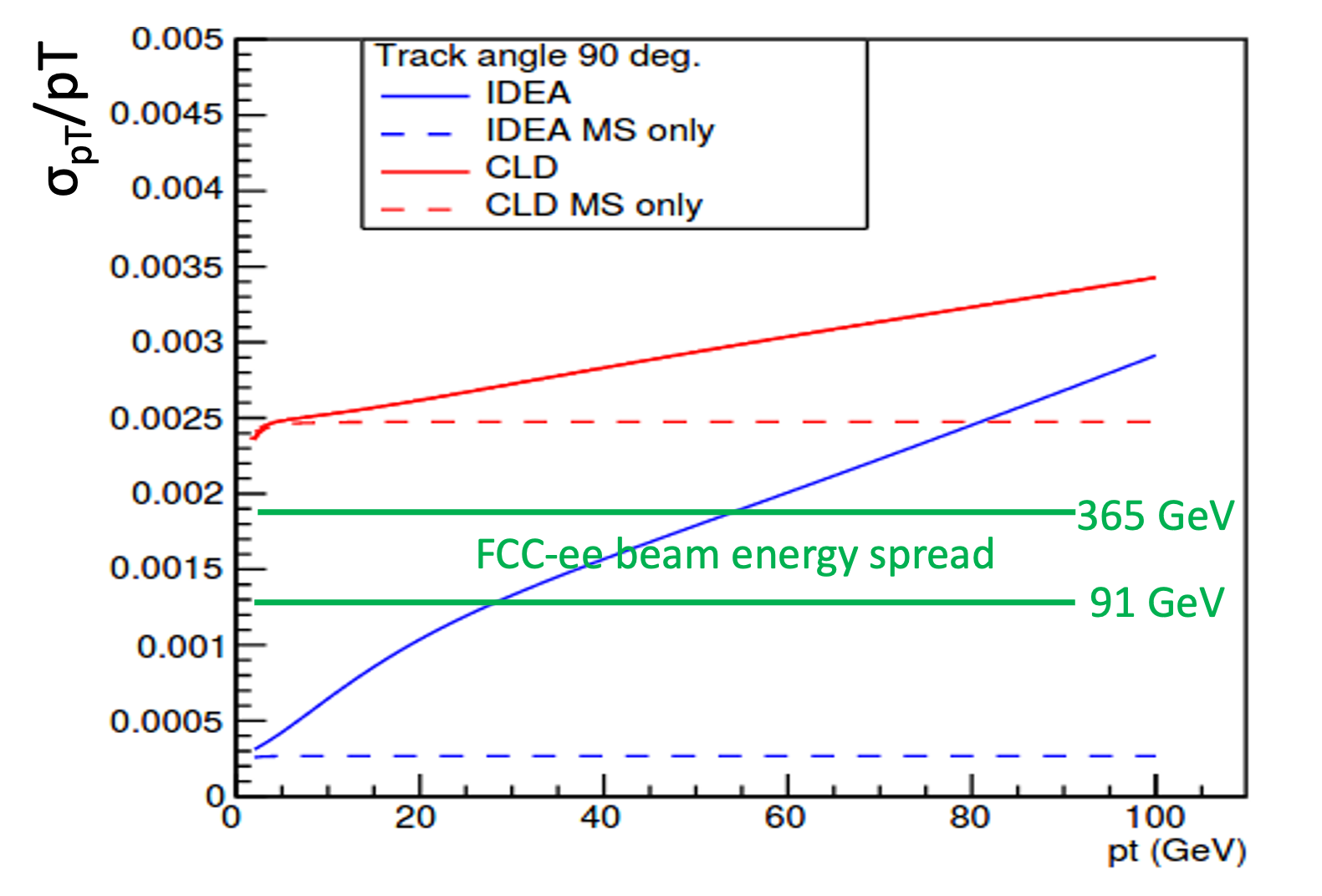}
\caption{Transverse momentum resolution as a function of $p_\mathrm{T}$, for tracks at $\theta = 90^\circ$, for two tracker technologies: in red, a full Si tracker as in CLD with a total material budget of 10\%; in blue, a tracker system as in IDEA with a highly transparent drift chamber surrounded by a Si wrapper. The contributions from multiple scattering are shown as dashed curves. As comparison, shown in green is the beam energy spread at $\sqrt{s} = 91$\,GeV and at $\sqrt{s} = 365$\,GeV.}
\label{fig:momres}
\end{figure}

An illustrative example of the difference in momentum resolution of the two tracker technologies is shown in Fig.\ \ref{fig:momres}. Resolutions are shown 
for a full Si tracker with a total material budget of 10\%, as exemplified by CLD, and for the IDEA tracking system comprising a vertex detector (2.2\% $X_0$), a highly transparent drift chamber (1.6\% $X_0$), and a Si wrapper. 
The better intrinsic point resolution of the Si-based tracker is reflected in the lower slope of the CLD curve compared to the IDEA curve. However, for the momentum range of interest, this advantage is 
overshadowed by the higher multiple-scattering contribution for CLD. The lesson learned is that, for FCC-ee, detector transparency is more important than point resolution; a strong case for gaseous trackers.


The CLD concept features an all-silicon tracker, comprising a vertex detector and a ``main tracker''. The vertex detector layout is based on double layers of MAPS (50\,\textmu{}m thick, $25 \times 25$\,\textmu{}m$^2$ pixels) fixed on a common support structure that includes cooling circuits. Located at radii below 112\,mm, the detector has three barrel layers and three disks in each direction. The main tracker is divided into inner and outer sections by a lightweight (1.25\%\,$X_0$) carbon-fibre support tube, at 690\,mm radius. The inner (outer) tracker comprises three (three) barrel layers and seven (four) forward disks in each direction (\mbox{Fig.\ \ref{fig:CLD}}). 
An engineering-level design is still to be performed; however, at the conceptual level, the material budget per layer, including sensors, electronics, cooling, and connectivity, is estimated to 1.09-1.28\%\,$X_0$ for the barrel and 1.28--2.13\%\,$X_0$ for the disks. 
%
In addition, carbon-fibre support structures 
contribute between 0.13\% and 0.37\%\,$X_0$
depending on the layer.
In total, the tracker material corresponds to about 10\%\,$X_0$ at 90$^\circ$ polar angle. Even if this is very transparent compared to trackers of the LHC experiments, it is clearly desirable to reduce the transparency further. Innovative solutions are needed.

The extremely light IDEA wire chamber, coaxial with the detector magnetic field, provides 112 space-point measurements, at radii between 35\,cm and 200\,cm, each with a resolution in $r\phi$ better than 100\,\textmu{}m.
About 350\,000 wires form 56\,448 approximately square, 12--14.5\,mm wide drift cells. 
Longitudinal information is provided via the
arrangement of the drift cells with 
alternating-sign stereo angles.
A very low material budget of $\sim$1.6\%\,$X_0$ at normal incidence (dominated by the outer wall) and $\sim$5.0\%\,$X_0$ in the forward direction (dominated by the end plates) is achieved via a novel approach for the wiring and assembly procedure~\cite{chiarello} and the use of a very light gas mixture, 90\% He + 10\% iC$_4$H$_{10}$. With a drift velocity of 2.2\,cm\,\textmu{}s$^{-1}$, the maximum drift time is $\sim$400\,ns. The number of ionisation clusters generated by a minimum ionising particle is 12.5\,cm$^{-1}$. Fast electronics, with a 2\,GHz sampling frequency, allows the cluster-counting/timing technique to be exploited, to improve both the spatial 
and the particle identification resolutions.

Recently, a straw chamber has been proposed as the main tracker. This detector has many similarities with the IDEA wire chamber; however, the drift cells are formed by cylindrical straws made with 12\,\textmu{}m-thick Mylar walls. A layout featuring $\mathcal{O}(60\,000)$ straws with a diameter of 10--15\,mm
arranged into 100 layers is being studied. With choices similar to those of IDEA for the gas mixture and the readout electronics, the space-point resolution and the particle identification capabilities are estimated to be similar. The material budget is $\sim$1.2\%\,$X_0$ at normal incidence, excluding the outer wall.

Time projection chambers (TPCs) provide continuous 3D tracking over a large volume with minimal material interference, while also enabling particle identification via 
energy deposition measurement in the gas. 
Ionisation electrons are drifted
to the detector endplates where the signal is amplified and registered. In recent R\&D efforts, endplate solutions based on Micro-Pattern Gaseous Detectors (MPGDs) with very 
high granularity (e.g., pixels down to $\mathcal{O}(100 \times 100\,\upmu{}\text{m}^2)$ size) have been studied. Important advantages 
include reduced ion backflow from the gas amplification process into the drift volume, and the ability to use cluster counting for improved PID performance. The potential operation of a TPC at the very high Z-pole luminosity is a subject under study. Backgrounds, related to beam-beam interactions, are the dominant source of space-charge build-up in the drift volume that must be minimised to reduce 
drift-path distortions.
To illustrate the challenge, it is instructive to consider the different typical time scales involved in the process: 
\emph{i})~25\,ns between BXs, 
\emph{ii})~5\,\textmu{}s between physcis events, 
\emph{iii)}~50\,\textmu{}s max drift time for electrons, and
\emph{iv)}~0.5\,s drift time for backflow ions. 
In other words, the detector has, at each moment of time, 
a memory of the previous $\mathcal{O}$(1000) BXs in terms of drift electrons, and of more than $10^7$ BXs in terms of ions, the latter demonstrating the importance of a very careful control of the ion backflow.


\subsection{Charged Particle Identification}

The identification of charged hadrons (\textpi{}, K, p) significantly enhances the physics potential of FCC-ee detectors. In particular, it improves the flavour-tagging performance in the selection of Higgs boson decays to $\mathrm{c\bar{c}}$ or
$\mathrm{s\bar{s}}$, and is crucial for the extensive flavour physcis programme.
The momentum region to be covered is up to a few tens of GeV.

\begin{figure}[ht]
\centering
\begin{subfigure}{0.495\linewidth}
\includegraphics[width=\linewidth]{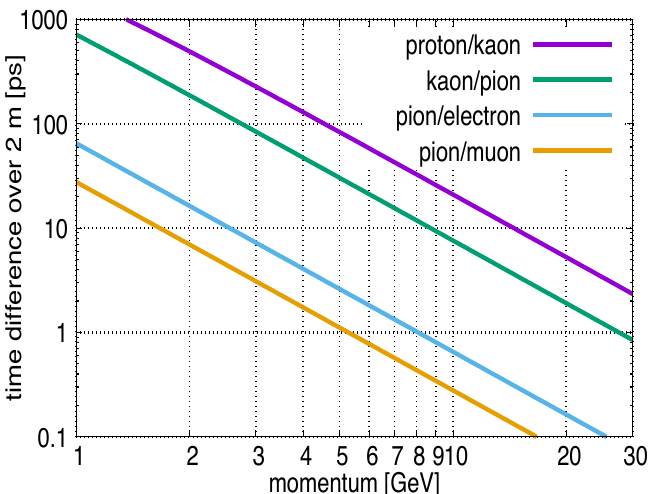}
    \caption{Time-of-flight differences}
    \label{fig:TOF}
\end{subfigure}
\hfill
\begin{subfigure}{0.46\linewidth}
\includegraphics[width=\linewidth]{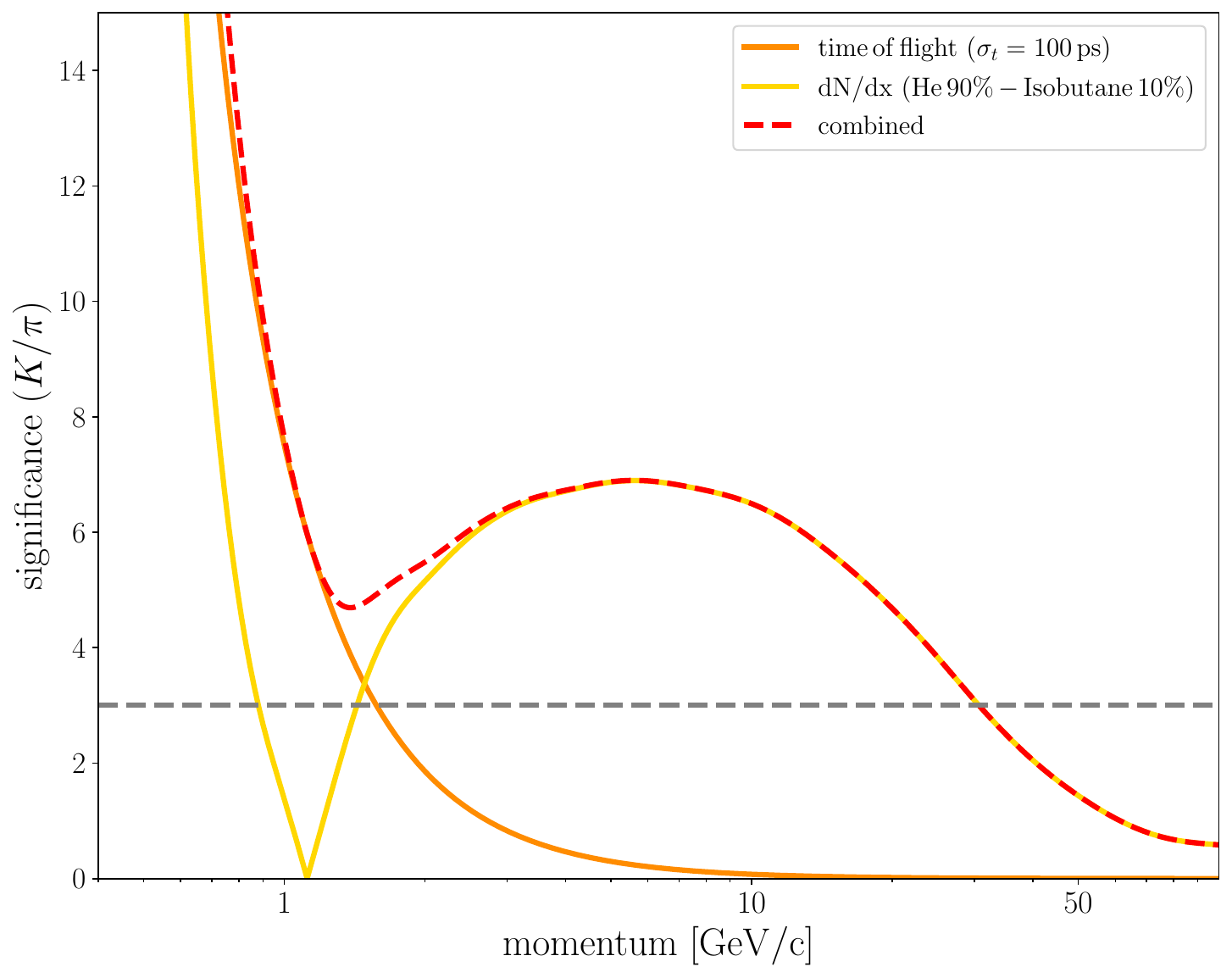}
    \caption{IDEA detector combined PID}
    \label{fig:IDEApid}
\end{subfigure}
\caption{Distributions illustrating particle identification capabilities.
(a)~Time-of-flight difference between particle types over a 2\,m flight distance as function of particle momentum.
(b)~Significance (in standard deviations) of the \textpi{}/K separation for the IDEA drift chamber (yellow curve).
The low momentum region is covered by a TOF measurement, over a distance of 2\,m and with resolution here assumed to be 100\,ps (orange curve). 
The combination of the drift chamber and the time-of-flight system is represented by the dashed red curve.
}
\label{fig:PID2}
\end{figure}

Time-of-flight (TOF) measurements, for example from a silicon-sensor layer 
surrounding the tracking volume, would provide discrimination between charged particles at low momenta. The flight-time difference between different particle types over a 2\,m distance is shown in Fig.\ \ref{fig:TOF}. 
While a 100\,ps resolution would provide a \textpi{}/K separation better than three standard deviations for momenta up to $\sim$1.6\,GeV, a 10\,ps resolution is required to extend the momentum range to $\sim$5\,GeV.

%

Gaseous trackers have been proven to provide good PID via the approach of correlating the momentum of charged particles with their specific energy loss, $\mathrm{d}E/\mathrm{d}x$. 
In fact, the number of ionisation clusters per track length, $\mathrm{d}N/\mathrm{d}x$, turns out to be an even stronger discriminant than $\mathrm{d}E/\mathrm{d}x$, which is hampered by the skew Landau distribution that governs the energy loss of the individual ionisation processes. 
The counting of clusters, separated by distances of $\sim$0.5\,mm, depends on a high sample granularity along the track, either spatially (for TPCs) or temporally (for drift chambers), recently 
made possible by novel read-out technologies. 
In a simulation study of the IDEA drift chamber 
the \textpi{}/K separation was found to be better than three standard deviations up to momenta of $\sim$30\,GeV, except in a narrow 0.9--1.6\,GeV gap, where the Bethe-Bloch energy-loss curves for the two particle types cross. 
The situation is illustrated in \mbox{Fig.\ \ref{fig:IDEApid}} where it is also demonstrated how the gap can be easily covered with a non-challenging time-of-flight measurement here assumed at a 100\,ps resolution.
A similar performance is obtained by TPCs with very high granularity readout.

\begin{figure}[ht]
\centering
\includegraphics[trim=0 -12mm 0 0,width=0.5\linewidth]{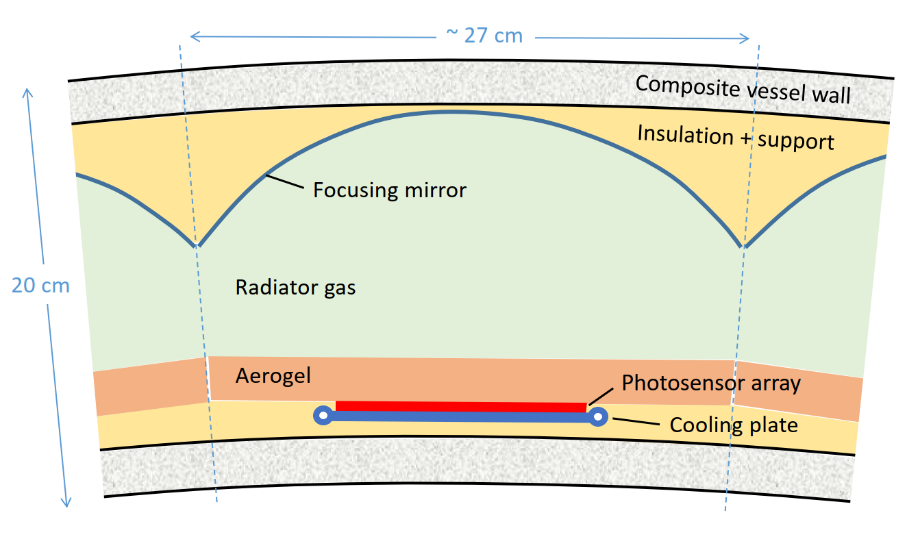}
\hfill
\includegraphics[width=0.48\linewidth]{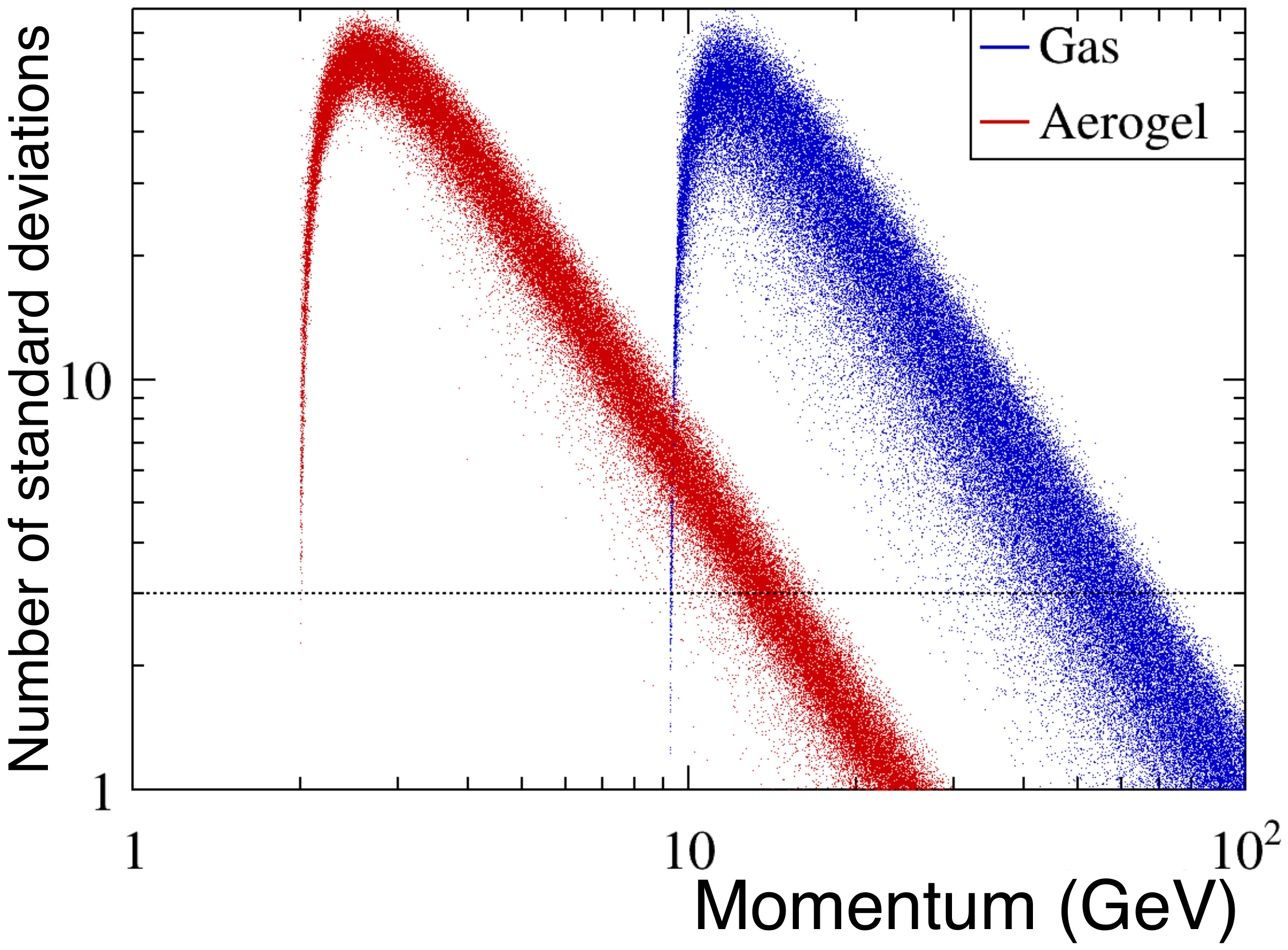}
\caption{ARC detector: Left: Schematic cross sectional view of the barrel detector, showing the various components of a single cell. Right: Momentum dependence of the K/\textpi{} separation, with two radiators.
}
\label{fig:ARC}
\end{figure}

For experiments with silicon-based tracking, where 
neither d$E$/d$x$ nor d$N$/d$x$ would 
provide the required level of 
performance, ring-imaging Cherenkov (RICH) detectors could deliver PID over a wide momentum range. 
Studies are underway to develop a compact RICH detector,
with the goal of fitting it in a 20\,cm radial envelope and contributing less than 10\%\,$X_0$ to the material budget, hence limiting its impact on the 
tracking and calorimetry. 
The compactness is enabled by use of
silicon photomultipliers (SiPM), 
that provide high detection efficiency and spatial granularity in a low thickness. 
A geometry has been developed, with a large number of similar RICH detector elements, in a cellular approach, a concept named ARC, for \emph{Array of RICH Cells}.
The components of an ARC cell are shown in Fig.~\ref{fig:ARC}. 
The lightweight vessel, made of carbon-fibre composite, contains a set of two radiators, silica aerogel and gas, with Cherenkov photons from both focused via a spherical mirror onto a common SiPM detector plane. The baseline choice for the gaseous radiator is C$_{4}$F$_{10}$ 
given its attractive optical properties.
With the high photon-detection efficiency achievable with SiPMs, 
this option provides a sufficient number of detected photons 
despite the limited radiator depth.
Given that fluorocarbons have a strong greenhouse effect, 
%
R\&D is required to find possible alternative gases, such as xenon with mild pressurisation, 
to ensure a sufficient photon yield. 
The aerogel radiator extends the PID performance down to momenta $\sim$2\,GeV. To cover lower momenta, a TOF system with a resolution of $\sim$60\,ps or better would be needed.

\subsection{Calorimetry}

Calorimeter systems are commonly divided into two separate compartments: the ECAL, used mainly to measure electrons and photons through their electromagnetic interactions, 
and the HCAL, used to measure mainly hadrons through their strong and electromagnetic interactions. This arrangement is also used for FCC-ee detector concepts, except for a variant of IDEA, where both electromagnetic and hadron showers are measured in a single, monolithic dual-readout (DR) calorimeter similar to the one used as the HCAL in the baseline layout.

The energy resolution of a calorimeter is generally written as
\begin{equation}
    \frac{\sigma(E)}{E} = 
    \frac{a}{\sqrt{E}} \oplus \frac{b}{E} \oplus c,
\end{equation}
where the first term is the stochastic term, and includes the shower intrinsic fluctuations; the second is the noise term; and the third is the constant term. At the relatively modest \mbox{FCC-ee} energies, an important role of the calorimeter system is to complement the tracking system in an optimal event reconstruction. To adequately classify multi-jet final states, a \mbox{3--4\%} two-jet invariant-mass resolution is needed to discriminate between \mbox{W-,} \mbox{Z-,} and Higgs-boson origins of two-jet systems. This again calls for a very good jet-energy resolution of $\sim$30\%/$\sqrt{E}$. Fine transverse and longitudinal segmentation facilitates an improved reconstruction of electromagnetic and hadronic objects, important for the so-called \emph{particle-flow} method, in which the energy of charged particles (typically $\sim$60\% of jet energies) is derived from track momentum measurements, the energy of photons ($\sim$30\%) is measured in the ECAL, and the energy of neutral hadrons ($\sim$10\%) is measured in the HCAL.
Optimal performance would occur in cases where the calorimeter object(s) from each incoming particle were completely separated from other objects and no object overlap appeared. 

The very broad FCC-ee physics programme, in particular that of flavour physics, poses a number of challenging requirements on the 
ECAL performance that accentuate those posed by the particle-flow method.
These include:
large dynamic range for measurement of photons between 100--200\,MeV and 180\,GeV; high granularity (and small Moli{\`e}re radius) for 
particle identification 
(e.g., electron/pion separation via shower-shape analysis, and separation of (collimated) decay channels such as $\uptau^\pm \rightarrow \uppi^\pm n \uppi^0$, $n=0,1$\ldots); 
very good energy resolution for 
particle identification (e.g., electron/pion separation via $E/p$ measurement) and for identification of exclusive heavy-flavoured-hadron decays with one or more final state \textpi$^0$s.

\begin{table*}[t]
\centering
\caption{Overview of calorimeter systems. Inspired from Ref.\ \cite{Aleksa:2021ztd}.}
\label{tab:ECALs}
\begin{tabular}{lccc} \hline
Detector Concept & \phantom{and ILD} CLD (and ILD) & IDEA & ALLEGRO \\ \hline
ECAL & & & \\
- Technology & W-Si & crystal & Pb-LAr \\
- Longitudinal segments & 40  & 2       & 11 \\
- Transverse granularity [mrad$^2$] & $2.5 \times 2.5$ & $5\times 5$ & $8 \times 10$ \\
- Energy resolution stocastic term & 16\% & 3\% & 8\% \\ \hline
HCAL & & & \\
- Technology & scintillator-steel & fibre-sampling dual readout & scintillator-steel \\
- Longitudinal segments & 44 & via timing & 13 \\
- Transverse granularity [mrad$^2$] & $10 \times 10$ & $1\times 1$ & $20 \times 20$ \\ \hline
Coil placement & outside calorimeters & between ECAL and HCAL & between ECAL and HCAL \\ \hline
\end{tabular}
\end{table*}
An overview of calorimeter systems under study is found in Table \ref{tab:ECALs}. For the ECALs, technologies include the extremely fine-grained tungsten-silicon (W-Si) sandwich structure of CLD and ILD; the high-granularity noble-liquid solution of ALLEGRO, a technology known to deliver excellent linearity, stability, and uniformity; and the finely segmented crystal layout of IDEA that provides dual readout capabilities where scintillation and Cherenkov signals are read out separately, for an improved energy resolution for hadrons. The electromagnetic energy resolution ranges from a relatively modest 16\%/$\sqrt{E}$ for the \mbox{W-Si} solution (due to a low sampling fraction), 
via 8\%/$\sqrt{E}$ for the noble-liquid Pb-LAr solution (6\%/$\sqrt{E}$ for a 35\% denser \mbox{W-LKr} solution) down to the excellent 3\%/$\sqrt{E}$ for the crystal solution. 
Two of the HCAL solutions (CLD and ALLEGRO) are based on steel absorber structures interleaved with scintillating tiles with SiPMs read-out.
The IDEA HCAL is based on fine structure of 
radially oriented fibres in a metal absorber read out by SiPMs at the outside. Dual readout capabilities are provided by the use of alternate rows of scintillating and Cherenkov fibres. 
With no segmentation in depth, 
studies are ongoing of 
how longitudinal shower information can be extracted from timing information.

An important difference between the 
detector concepts is the placement of the superconducting coil that provides the detector magnetic field. In order not to disrupt the very detailed shower imaging capabilities of their calorimeter system, CLD and ILD have 
the coil outside the HCAL. For both IDEA and ALLEGRO, the coil is placed between the ECAL and HCAL, and it becomes relevant that the coil, including cryostat, is as thin and transparent as possible. It has been estimated~\cite{fcc-ee-cdr} that a system of coil and cryostat can be constructed within a radial envelope of $\sim$30\,cm and with a material budget of 
about 0.8\,$X_0$.

\subsection{Muon System}

Muon systems are implemented inside gaps in the magnet return yoke.
Apart from muon identification, these systems offer capabilities for tail catching of calorimeter showers and identification and measurement of long-lived particles by providing tracking with good position resolution. Several technologies are considered, including scintillators, drift tubes, resistive plate chambers, resistive micromegas, and \textmu{}-RWELLs.

\subsection{Luminosity measurement}

For precise cross-section measurements, the integrated luminosity must be determined with high precision. 
Ambitious goals have been formulated on the measurement precision of
$10^{-4}$ or better on the \emph{absolute} luminosity
and about $10^{-5}$ on the \emph{relative} luminosity between energy scan points. 
As a complement to the traditional small-angle Bhabha scattering process, e$^+$e$^- \rightarrow$ e$^+$e$^-$, the wide-angle diphoton process, e$^+$e$^- \rightarrow \upgamma\upgamma$, is statistically relevant and may allow moving beyond the 10$^{-4}$ absolute luminosity goal.
%
For both processes, the definition of the geometrical acceptance constitutes an important source of systematic uncertainty.
More details can be found in Refs.\ \cite{Dam:2021sdj,blondel_2023_f1fs5-0jr59}.

For normalisation via the diphoton process, a minimum scattering angle $\theta^{\upgamma\upgamma}_\mathrm{min} \simeq 15^\circ$ is relevant for the acceptance definition.
At this angle, a $2\times 10^{-5}$ relative uncertainty on the acceptance corresponds to an uncertainty $\delta \theta^{\upgamma\upgamma}_\mathrm{min} \simeq 8$\,\textmu{}rad,  
corresponding to a 20\,\textmu{}m tolerance on the radial coordinate at a distance of 2.5\,m from the IP where photons are measured in the forward ECAL.

For measurement of small-angle Bhabha scattering, the LumiCals are 
placed around the beam pipe neighbouring
the complex MDI region, \mbox{Fig.\ \ref{fig:vtx}}. 
Located in front of the compensating solenoids, the LumiCals are at only $\sim$1.1\,m from the IP. 
In this region, space is severely limited, and a very compact detector design is necessary. A solution based on a dense W-Si sandwich is under study.
The Bhabha-scattering cross section is very steeply dependent, $1/\theta^3$, on the scattering angle $\theta$.
This results in a very challenging tolerance on the inner radius of the acceptance at the $\mathcal{O}$(1\,\textmu{}m) level.

\subsection{Trigger and data-acquisition system}

The target of the data-acquisition system is to be fully efficient to all Standard Model 
annihilation events and not to miss signatures of possible BSM phenomena, such as, for example, heavy, slow-moving, late-decaying particles, LLPs. Two possible architectures can be considered: \emph{i}) a \emph{triggered} system (as known from LEP and LHC experiments), where prompt data from a subset of detector components are used to derive a fast decision which, if positive, triggers readout of the complete detector; and \emph{ii}) a \emph{trigger-less}, \emph{free streaming} system (as being implemented for the ePIC experiment at EIC \cite{VCI-EIC}) where each subdetector autonomously pushes out all observed hit data to an off-detector event-builder system where data segments are assembled and events are filtered. For both possibilities, as in general, the largest challenge arises from the huge Z-pole event rate.

Conceptually, a trigger system would be similar to the systems employed at LEP, where the physics target was the same \cite{LEPtrigger2015}. By the very minimal requirement of observing at least a single charged or neutral particle, trigger efficiencies for hadronic Z decays of 
($99.999\pm 0.001$)\% 
were observed. Typically,
trigger rates exceeded the rate of Z decays by 
an order of magnitude.
Extrapolating this to FCC-ee would result in a trigger rate of 500\,kHz, approximately 1/100th of the BX rate. Technically, however, a trigger system would be more like the systems known from LHC experiments. With a similar high BX rate, it would be necessary to save detector data locally in so-called latency buffers awaiting the
decision of the trigger system. Such local buffering risks being expensive in terms of material budget and power consumption and could compromise 
the detector resolution.

For a trigger-less solution, a bottleneck is likely to arise from the transport of very large data volumes away from detector elements. 
The most challenging is probably the vertex detector where,
according to simulation studies, the inner layer will see a rate of $\sim$15\,MHz\,cm$^{-2}$ of background particles from beam-beam interactions. With an assumed cluster size of five and a safety factor of three, the pixel hit rate will then be $\sim$200\,MHz\,cm$^{-2}$. With, say, 24 bits per pixel (address + signal height) this corresponds to a data rate of $\sim$5\,Gbit\,cm$^{-2}$\,s$^{-1}$. Studies are needed to assess whether such a local data rate would be possible. In any case, the total data rate out of the detector would be very large, likely reaching several tens of GBit per second.

\section{Outlook}

%
In the 2020 update of the European Strategy for Particle Physics (ESPP) it was concluded that ``\textsl{An electron-positron Higgs factory is the highest-priority next collider. For the longer term, the European particle physics community has the ambition to operate a proton-proton collider at the highest achievable energy}''. As a reaction, CERN, in 2021, launched the FCC Feasibility Study. 
The study results have recently been made publicly available 
through the FCC Feasibility Study Report~\cite{Benedikt:2025hsi,Benedikt:2928793,Benedikt:FSRv3}, as input to the ongoing 2026 ESPP update process. 
Given a positive ESPP recommendation 
and a subsequent timely approval by the CERN Council, the FCC construction would proceed.
For the detector design effort, resources can be expected to increase significantly over the next 3-5 year period following the finalisation of the high-luminosity upgrades of the LHC detectors. This would allow the design and construction of four ambitious FCC-ee detectors 
over the next decade for installation in the first half of the 2040's.



\section*{Acknowledgements}

I would like to thank the organisers of VCI2025 for giving me the opportunity to present this work in the inspiring environment of their conference.

\bibliographystyle{elsarticle-num}
\bibliography{vci2025.bib}

\end{document}